\RequirePackage{fix-cm}
\documentclass[smallextended]{svjour3}       
\smartqed  
\usepackage{graphicx}
\titlerunning{Higher Theory and the Three Problems of Physics}        
\institute{
\email{a.veilahti@gold.ac.uk}   
}

\date{\today}

\usepackage[T1]{fontenc}   
\usepackage{lscape}

\usepackage[all,cmtip,2cell]{xy}
\UseTwocells
\usepackage{ucs} 
\usepackage[applemac]{inputenc}
\usepackage{textcomp}  
\usepackage{t1enc}    

\usepackage{graphicx}
\usepackage{adjustbox}

\usepackage{ae}
\usepackage{amsfonts}         
\usepackage{amsmath}
\usepackage{amssymb}
\usepackage{setspace}
\usepackage{upgreek}
\usepackage[T3,T1]{fontenc}
\DeclareSymbolFont{tipa}{T3}{cmr}{m}{n}
\DeclareMathAccent{\vbar}{\mathalpha}{tipa}{156}

 	\newcommand{\R}{\mathbb{R}} 
\newcommand{\C}{\mathbb{C}}

\newcommand{\E}{\mathcal{E}}

\renewcommand{\P}{\mathbb{P}}

\newcommand{\Hom}{\textrm{Hom}}

\newcommand{\End}{\mathrm{End}}

\newcommand{\Bt}{\mathfrak{B}}

\newcommand{\As}{\mathscr{A}}

\newcommand{\Spec}{\textrm{Spec }}

\newcommand{\Sys}{\mathpzc{S}\mathpzc{ys}}

\newcommand{\F}{\mathcal{F}}

\newcommand{\Hs}{\mathscr{H}}

\DeclareMathAlphabet{\mathpzc}{OT1}{pzc}{m}{it}
\newcommand{\sets}{\mathscr{S}\mathpzc{ets}}
\newcommand{\A}{\mathcal{A}}

\newcommand{\B}{\mathcal{B}}

\newcommand{\V}{\mathscr{V}}
\usepackage{fge}

\renewcommand{\Xi}{\Upsigma}

\newcommand{\Lie}{\mathrm{Lie}}
\newcommand{\Gs}{\mathscr{G}}
\newcommand{\Det}{\mathrm{Det}}

\newcommand{\Ns}{\mathscr{N}}

\usepackage{slashed} 

\newcommand{\G}{\mathcal{G}}
 	\newcommand{\Ell}{\mathcal{L}}
 	
\newcommand{\Cat}{\mathpzc{C}\mathpzc{ats}}

\newcommand{\Loc}{\mathscr{L}\mathpzc{oc}}

\newcommand{\Grp}{\mathcal{G}}

\def\F{\mathcal{F}}
\newcommand{\Rep}{\mathcal{R}ep}

\usepackage[mathcal]{eucal}
\usepackage{eufrak}
\usepackage{mathrsfs}

\usepackage{etoolbox}

\makeatletter
\patchcmd{\@settitle}{\uppercasenonmath\@title}{}{}{}
\patchcmd{\@setauthors}{\MakeUppercase}{}{}{}
\makeatother

\usepackage{fge}

\author{Antti Veilahti}

\title{\Large{Higher Theory and the Three Problems of Physics}}
\author{Antti Veilahti}

\begin{document}

\thispagestyle{empty}

\maketitle	
\pagenumbering{roman}

\noindent \textbf{Abstract} According to the Butterfield--Isham proposal, to understand quantum gravity we must revise the way we view the universe of mathematics.
However, this paper demonstrates that the current elaborations of this programme neglect quantum interactions.
The paper then introduces the Faddeev--Mickelsson anomaly which obstructs the renormalization of Yang--Mills theory, suggesting that to theorise on many-particle systems requires a many-topos view of mathematics itself: higher theory. As our main contribution, the topos theoretic framework is used to conceptualise the fact that there are principally three different quantisation problems, the differences of which have been ignored not just by topos physicists but by most philosophers of science. We further argue that if higher theory proves out to be necessary for understanding quantum gravity, its implications to philosophy will be foundational: higher theory challenges the propositional concept of truth and thus the very meaning of theorising in science.






\thispagestyle{empty}

\pagenumbering{arabic}

$\ $

\noindent \textbf{Keywords}

\noindent Topos quantum physics, category theory, higher theory, quantum interaction, quantum gravity, Yang--Mills theory

\tableofcontents

\thispagestyle{empty}

\pagenumbering{arabic}

\setcounter{footnote}{0}
\section*{Introduction}
As one of the  Millennium Prize problems set by the Clay Institute, Yang--Mills theory is amongst the most puzzling dilemmas of contemporary science. Though it might appear as a modest concern that ‘the quantum particles have positive masses, even though the classical waves travel at the speed of light'\footnote{For a full definition, see \\
http://www.claymath.org/millennium-problems/yang-mills-and-mass-gap.}, this problem surrounds something much deeper. 
Not only there is a conflict between the theories of physics at small and large scales (or interaction and gravity). Instead, we argue, the ‘mass gap' mirrors the very distance between physical phenomena and their correlates in mathematical physics. 

This paper contextualises this gap from the point of view of \textit{topos theory}---a still little known area of contemporary mathematics that, nevertheless, bears broad implications to both philosophy and science. Since Aristotle, it has been foundational to Western thinking that propositions are divided into true and false statements. It is this very principle that is now changing: by deriving from Paul Cohen's [9, 10] work on the shortcomings of set theory, topos theory applies geometric methods to the concept of truth.

This is both a radical and challenging idea, not least because the erudition in mathematics it requires. Topos theory is based particularly on the so-called ‘categorical' methods developed in the latter half of the twentieth century. They are central to many fields of mathematics (e.g., the proof of the Fermat's last theorem), yet only limited literature in either philosophy or physics  touches them [e.g., 2, 32, 37, 48]. And even if the true contributions of topos theory are still open to dispute, Cohen's discovery of the incompleteness of set theory is generally accepted as a relevant turning point in the history of formal logic.

Illustrating the pertinence of these concerns to physics, research programmes like the \textit{Deep Beauty} suggest that the entire universe of mathematics needs to be replaced to understand  quantum gravity [24]. There is a need for ‘a fundamentally new way of constructing theories of physics' [14]. Andreas Döring and Chris J. Isham [13, 14, 15] themselves attempt to develop a topos ‘inside' of which the laws of physics could be expressed. In other words, an elementary topos is often said to be a ‘universe of construction' where mathematics (e.g., set theory) ‘is possible'\footnote{This means that set-theoretic statements can be emulated on the level of diagrams related to the subobject classifier $\Omega$ of that topos. These diagrams are then used to express propositions, theories and proofs.} [30]. Topos quantum physics could thus be viewed as a ‘reflexive' approach to mathematical physics, allowing us to emulate mathematically the emergence of philosophically important concepts like quantisation and materiality. 

Yet a notable degree of ambiguity remains regarding the meaning of the proposal Isham introduced first with Butterfield: it could stand for very different things depending on in what particular kind of a topos such theories are being constructed.  Döring and Isham are right in that the topos in which physical theories take place needs not be the ‘classical' universe where the Zermelo--Fraenkel axiomatic holds. But the so-called ‘elementary' topos theory that envelopes Döring's and Isham's  argument is only the first example of an alternative universe of mathematics.

This paper argues that not just the current approaches to topos physics but also the philosophy of physics is  most often guided by what we will call the ‘first quantisation' problem---the one dominated by the Einstein--Bohr-debate [e.g., 6]. Like Pickering [43] argues, it is based on different questions than the development of field theory, that is, the second generation of quantum theories that became dominant in the 1970's. For instance, philosophers tend to restrict to the ‘quantum foundations' problem whenever discussing the relevance of category theory to physics.  Similar tendencies overshadow topos physics where no topos-theoretically erudite accounts besides Urs Schreiber's [46] work address interaction and gravity at the level of their actual mathematical contents. 

It is not the main task of this paper to contribute technically to these fields.  Instead, following a different philosophical schematic, we examine, in a mathematically erudite way, the \textit{conditions} of theorising on such foundations. Two main goals follow.

First, we argue that there are \textit{three generations} of quantum theory: three kinds of empirical problems each associated with their own approaches to mathematical reasoning. We conceptualise them by introducing a brief genealogy of some---but, given the limited space, in no way all---theories pertinent to the three quantisation problems. They should be illustrative also to those philosophers of science who lack previous familiarity with quantum physics (particularly its second problem). 

This leads us to another challenge, which relates to the modes of reasoning we engage. Indeed, as the second goal, we argue that the ‘reflexive' aspects of  topos theory  provide a way to mathematically examine  the structures of reasoning behind some of the most noble quantum philosophical dilemmas. This is only possible by allowing topos theory to inform philosophical reasoning. 
For instance, the topos perspective allows us to specify the exact relationship of the quantisation problems and  the idea that physical theory should emerge as an ‘internal' part of the world, that is, that there is no need to posit an external, artificial observer. 

In effect, if topos physics provides a ‘reflexive' approach to the three generations of quantum theory---as our main argument goes---the philosophy of physics should be organised accordingly.  Most philosophers of science, however, subscribe to a very different kind of paradigm, viewing formal logic as constitutive to philosophical argumentation. From the topos perspective, it is precise status of these rules of logic that has changed---if they exist at all. 

To summarise, after demonstrating that the current, ‘elementary' approaches to topos quantum physics are limited, we will ask whether the use of higher categories help us understand something that set theory does not. In particular, where elementary topos theory emulates the structures of set-theoretic language (considered as 0-geometric), there is a higher categorical analogue developed by the $\$3$ million \textit{Breakthrough Prize} nominee Jacob Lurie [36], and it emulates more complex,  ‘$n$-geometrical' objects instead (Table 1). What is particularly interesting is that a higher topos can be viewed as a combination of multiple lower degree topoi, and as we will demonstrate in the context of Faddeev-Mickelsson anomaly, this is connected to the way in which physics approaches systems combining multiple particles. From this point of view,  mathematics is not only used to ‘describe' physical systems but it folds around them. 

\begin{table}[t]
\label{evolution}
\caption{
Evolution of topos theory and its connections to physics. }
\begin{adjustbox}{width=1\textwidth}
\begin{tabular}{rcccccc}
	& & local theory  & elementary theory & higher theory  & \ldots & higher theory \\
theory && 0-topos & 1-topos & 2-topos & & $\infty$-topos \\ \ \\
\hline
\ \\
external objects && sets & sheaves &  stacks & \ldots & $(\infty, 1)$-categories \\
&&&& fibred categories &\\
internal appearance && set theory & local theory & elementary theory & & higher theory \\ \ \\
description && non-categorical & 1-categorical & 2-categorical & & higher categorical \\ \ \\
classifying object & & true/false & $\textrm{true} \subset \Omega$ &  \multicolumn{3}{c}{full and faithful opfibration $\Omega_* \to \Omega$} \\ \ \\
\hline \ \\
an example && von Neumann topos & torsor $BG$ & gauge classification &  \ldots &  homotopy categories  \\ \ \\
relevance to physics & & quantum mechanics & reflexive theory & field theory & & quantum gravity \\ &&&&& &final theory?
\end{tabular}
\end{adjustbox}
\end{table}

The argument comprises the following three steps. First, we argue that the philosophy of quantum physics, together with the existing approaches to topos quantum physics, tends to ignore the relevance of the shift to the second quantisation problematic.
Second, the paper illustrates the relevance of higher topos theory, by deriving from its ability to conceptualise many-body systems mathematically and by arguing that higher theoretical point of view is thus required for understanding quantum interaction. 
Third, the paper argues that topos theory allows us to address one of the most challenging philosophical dilemmas: that quantum theory---the ‘subject’ seeking to observe quantum phenomena---is in a reciprocal relationship with the physical reality itself. The next section provides an overview of this argument, keeping particularly those readers in mind who find themselves less preoccupied of working through the mathematical contents (Sections 2--5).

\section{Overview of the Argument}

What separates quantum physics from classical physics relates to its concern for the question of \textit{quantisation}: it does not presuppose matter as being divided into separate particles, nor ignore them, but,  rather, it seeks to explain how things \textit{become} quantifiable. 
In other words, instead of starting with a system of separate particles, quantum science starts from a dynamical system from which individual particles emerge only in retrospect. Measures like momentum and position can then be viewed as operators on the solutions of such a system, that is, on the so-called wave-functions.

We refer to this general interest of quantum science as the family of \textit{quantisation problems}: how quantities are presented through material phenomena. Of course, 
any single, technical approach to the quantisation problem involves an attempt to ‘renormalise' this picture, as physicists call it, and thus replace ‘material phenomena' by another set of mathematical entities. 
Quantum mechanics, for example, starts with the Scrhödinger equation. Even if the infinite dimensional Hilbert space then reveals itself in a finite dimensional form---something quantifiable even to the classical intuition---this means that quantities do not ‘emerge' from material phenomena themselves but, instead, from the state space that replicates the latter. The individuation of particles is still theoretically rather than empirically informed. 

Whatever the case, in the context of the first quantisation problem, Döring and Isham carefully examine the conditions under which such a quantisation procedure could be posed even from the mathematical perspective. Under what topos theoretic conditions does the classical interpretation become intelligible? 
In particular, is the exact topos within which quantum phenomena (e.g., the state space) should be articulated irreducible to the latter? These questions replicate the question of emergence in a twisted way: we ask whether quantum reality, that is, the ‘topos' in which theory is articulated, is itself  irreducible or contingent in relation to quantum phenomena.

At least some level of contingency seems to exist, as we will discuss, because there turns out to be two different approaches to elementary topos quantum physics: Einstein's objectivist and Bohr's subjectivist interpretations can be supported by the same phenomena. Yet the scope of the emergence-problematic is still limited in this context: there appears to emerge only a single choice about the reality as a whole (subjectivist/objectivist views over quantum topos), whereas many-body systems require a situated, local perspective on emergence. In result, the early approaches to topos quantum physics are limited in that they apply only in the context of systems consisting of just a \textit{single particle}. Quantum mechanics is ‘not yet the real thing', as Einstein's ‘inner voice' told him in 1926. 

But how could the question of quantisation then be posed in any other, more ‘local'  way? The new developments of physics since the 1950's have largely evolved around this question. The second quantisation problem, which we will explain in more detail in Section 3, \textit{localises} the idea of quantities and invariance, that is, the asymmetry inherent to the renormalisation problem. And topos quantum physics, we argue, similarly localises the question of mathematical invariance, although not in respect to the ‘contents' of quantum theory but as relative to how it is being ‘expressed'. 
We claim that an exhaustive resolution of the mass gap -problem would entail the convergence of these two framings of the quantisation problem. 

To understand the prospects of discovering such convergences, we will first demonstrate that quantum mechanics still assumes mathematical invariances (like quantum states) to ‘globally' exist. This is equivalent with the concept of invariance or equality in elementary topos theory, where truth-values are assumed to satisfy a global hierarchy---a world whose internal form is ‘essentially propositional', like the one of Wittgenstein's \textit{Tractatus}. This assumption is implied by the two existing approaches to topos quantum physics. Both approaches are then limited in two ways: at the level of expression, they only employ 1-categorical means (elementary theory), which means that the phenomena ‘internal' to such theories can only resonate with single-particle systems.
Moreover, both approaches restrict to \textit{local} topos theory. Despite Butterfield's and Isham's original proposal, this means that their own concept  of mathematical invariance is still classical, and their attempt to theorise on many-topos systems hardly results in anything groundbreaking. 

Theories of quantum interaction, by contrast, \textit{localise} the very idea of mathematical \textit{invariance} (the so-called phase invariances, see Section 3). It would be artificial to assume a measuring device to exist outside quantum reality, but measurements are instead made through interactions with other particles, and the ‘gauges' are thus considered as local or situated, like Yang and Mills [54] suggested in 1954. 

Likewise, in Section 5 we will argue that topos theory provides an analogous, situated view on numbers and measurements. In particular, 
we will demonstrate that there is an analogue between the notions of ‘locality' in the technical sense of Yang--Mills theory, and as conceptualised by topos theory.  Yang--Mills theory is, of course, only one (conformal) approach to quantum interaction, but as a matter of fact categorical tools are now applied also in the context of non-conformal field theories like string theory [47].  We will then suggest that the many-body systems of physics could be viewed as being mirrored in many-topos systems of mathematics. Quantum mechanics, by contrast, is articulable in a non-situated way only because it assumes the independence of the concept of measurement in the first place.

This analogue holds the key to understanding what is at stake in contemporary quantum science: that we need an \textit{interactive} view of not only quantum phenomena but of mathematics as well. 
The resulting ‘$n$-categorical' framing of physics [cf. 3] are particularly eminent because they suggest that we should give up the very language of logical descriptions, but approach truth and reality geometrically instead. In particular, this perspective challenges us to ask whether Arthur Jaffe and Edward Witten's\footnote{First, ‘every excitation of the vacuum has energy at least $\Delta$'---or that the Hamiltonian $H$ has no spectrum in the interval $(0, \Delta)$. Furthermore, despite the fact that some quark fields might transforms non-trivially under $SU(3)$, elementary particles should anyway submit to such symmetries according to ‘quantum confinement'. Finally, in accordance with the ‘chiral symmetry breaking' the vacuum itself is invariant only under a particular subgroup of the full symmetry group. The problem is now to prove that  for a compact, simple gauge group $G$ a non-trivial quantum Yang--Mills theory exists on $\R^4$ and has a mass gap $\Delta > 0$. [29.]} set-theoretic framing of the ‘mass gap' problem is an adequate basis for understanding it.

There is also a philosophical demand for more elaborate interpretations. Unfortunately, most philosophers of physics have little insight into the new directions---propositional logic still dominates even the areas  focusing specifically on quantum gravity [e.g., 11, 40, 38], the ‘elusive' Higgs mechanism [49], or string theory [cf., 4]. While empirically discovered only in 2012, this ‘elusive' mechanism was proposed already in 1962, though first rejected by \textit{Physics Letters} as being ‘of no obvious relevance to physics'. Without it, the second quantisation problem of  many-body systems would have been only a ‘warm-up exercise' [49]. 

Nevertheless, even if the world itself is no longer overshadowed by what Huw Price [44] metaphorises as the ‘block universe view', mathematics itself is anyway viewed exactly as such a logically partitioned universe---the so-called local topos of sets. Given how paradigmatic role formal still plays in the philosophy of science, many physicists themselves now draw from much older intuitions, including Hegel [45, 55], Heidegger and Kant [15: 755, 800]. 

As a way of exemplifying this philosophical shortage, the obstruction to the second renormalisation problem of Yang--Mills theory is traditionally pronounced by the Coleman--Mandula Theorem. It states that the global symmetries of the so-called Poincaré-group are incompatible with local dilation invariance [see 41]. The standard resolution is to point out a loop hole to that theorem: it assumes the spacetime to exist before that full Lie group symmetry is broken [e.g., 35: 60--61]. This, however, is to suggest that the ‘historical' (breaking) is not the same as the ‘mathematical' (symmetry), bordering on similar distinctions dominating speculative realism in philosophy. The argument falls short because the gap between the two is preconceived and is neither theoretically nor empirically validated. Higher topos theory, we will argue, provides another interpretation as a situated, and thus a ‘historical' view of mathematics itself.

This study frames this possibility from the point of view of the three quantisation problems of physics---quantum mechanics, field theory and loop quantum gravity. It is reasonable to consider these as different approaches to the problem of temporality which, following Kant, can be reflected by three different phrases: ‘success', ‘coexistence', and ‘permanence'. Let us next operationalise these concepts mathematically (Sections 2--5) before discussing their implications for the nature of theorising in physics (Section 6). 


\section{First Quantisation---The World Internal to a Quantum}
\label{first}

The first quantisation problem, which is the basis of classical quantum mechanics,  can be viewed as a \textit{single} particle system -based reflection of \textit{Heisenberg's uncertainty principle}. Kennard and Weyl then formulated the principle as the inequality  $\sigma_x \sigma_p \geq \frac{\hbar}{2}$ based on two operators:  momentum and position. One of the two could only be known on the detriment of the other. However, some suitable combinations of quantum operators---the so-called self-adjoint operators---still carry a determinate value: one of them is the total energy operator $H$ (Hamiltonian), which gives rise to the Schrödinger-equation
\begin{equation}\label{Scrhö?dinger2}i\hbar\frac{\partial\psi}{\partial t} \psi(t, x) = H(x, -i\hbar\nabla)\psi(t, x).
\end{equation} A given solution, a so-called \textit{quantum state} $|\psi\rangle$, then appeared to hold qualities similar to probabilistic density functions. This was consequentially used as a basis for assessing the probabilistic likelihood of discovering a given particle in a specific domain of spacetime. 

A quantum state is, of course, still only a ‘synthetic' construct: individual ‘states' could only collapse upon empirical observation. However, by speculating on the existence of all such states, it was possible to \textit{represent} physical operations as operators on the state space as a whole\footnote{ 
Mathematically, quantum operators were traditionally defined as bounded operators closed under weak operator topology resulting in a so called $W^*$-algebra equipped with a Hermitean structure that extends complex conjugation. Determinate ones, like the Lagrangian regulator, would then be \textit{self-adjoint} ($A^*A =1$), that is, unaffected by the ‘subjective' order in which they were applied.}. What is striking is the recognition of the fact that the quantum reality is not independent of the \textit{order} in which measurements like position and momentum are recorded (referring to the first mode of temporality).

$\ $ \\ \noindent\textit{2.1 Döring's and Isham's Approach to ‘Daseinisation'}

\vspace{.2cm}

Döring and Isham, almost a century later, address this classical interpretation  \textit{categorically}, by the means of elementary topos theory. Instead of restricting to self-adjoint operators which globally commute (and are considered ‘determinate’), they focus instead on commutative subalgebras. The determinist interpretation is then possible \textit{inside} but not across contexts: the order in which operators are applied is redundant only if they come from a shared context. 

Such commutative contexts, in turn, form a lattice that gives rise to the so-called \textit{von Neumann topos} consisting of ‘sheaves' over that lattice $\V(\Ns)$. The approach is never less intriguing as Döring and Isham [14] demonstrate how mixed quantum states require the quantum topos to be extended by the probability topos, thus using the illustrating the probabilistic interpretation as the example of how the topos perspective can be used to address uncertainty. 

However, to illustrate the true  relevance of topos theory, each topos actually combines two different situations: ‘internal' and ‘external'. The internal situation, in Döring and Isham's elementary case, consists of a language similar to traditional quantum algebra. Yet the ‘external' status of that language is another, topos-theoretic problem. The classical ‘paradoxes' of quantum mechanics appear to stem from the (mis)conception assuming an equivalence between the ‘internal' and ‘external' situations of truth (this assumption is correct only in \textit{local} topos theory).

Formally, the quantisation problematic internal to a topos can be expressed as an arrow $\underline{\check \Xi } \to \underline{\check \R}$. We could say that this is the arrow which makes the question of ‘quantisation' possible. In the set-theoretic topos, for instance, it is used to grade different quantum states according to real numbers. Döring and Isham reflect it as the ‘daseinisation' of a physical system. Indeed, the ‘daseinisation' allows classical propositions of the form ‘$\Lambda \in \Delta \subset \R$' to be studied as ‘things' accessible from inside the von Neumann topos\footnote{Even if the so-called spectral presheaf $\underline{\Xi}$ itself is not a locale---it does not even have a single global element according to the Kochen--Specker theorem---Döring and Isham introduce a subsequent construction of a locale $\underline{\check \Xi}$ associated with the \textit{closed open topology} on $\underline{\Xi}$. It materialises the ‘state locale' \textit{internally}  to the von Neumann topos, resulting in a language regulating physical phenomena relative to that object $\underline{\check \Xi}$---a body of a quantum ‘internal' to the site.}. 

By internalising the relationship between physical phenomena and the rule of law, the state locale then serves as an intellectual ‘boundary object' between mathematics and physics (to employ a phrase typical to science and technology studies [7]). It gives rise to a peculiar language while, at the same time, it materialises the semantic of that language.

However, there are two restrictions to Döring's and Isham's discovery of such an  an object. First, 
Döring and Isham consider the universe only as a \textit{local topos}. It is not only that the concept of propositional language limits the ‘interiority' of the phenomena, like in the broader class of elementary topoi, but also the ‘external' truth is limited in a way ‘local' in respect to set theory (the lattice $\V(\Ns)$ exists as a set). Topos theorists call such topoi as being ‘logically bounded' [see 30: 150]. 

To Döring's and Isham's defense, of course, the von Neumann topos is only the first example of a quantum topos. However, there are also conceptual reasons to assume that local theory plays a peculiar role in their thinking. Namely, even if Döring and Isham [15: 912] acknowledge that ‘quantum theory can be viewed [$\ldots$] in a topos other than' $\sets$ (where the internal axiom of choice not necessarily applies), their discussion of ‘the category of systems' $\Sys$ [15: 871] is reasonable only in the context of local topos theory. We could thus say that only because they restrict to local topos theory, they can avoid encountering the problem of many-body systems and quantum interaction, hoping to resolve the ‘many-world'-dilemma in a way still regulated by a global lattice of truth-values, that is, the category of all locales $\Loc$. 

As a second concern pertaining to the problem of interaction, the theory incorporates  ‘daseinisation' only as a single, global arrow. Therefore, even if the existential decision to restrict to local topoi was abandoned, an elementary approach to topos quantum physics, too, would fail to incorporate the so-called  natural equivalences (i.e. ‘symmetries') of this arrow. In effect, even if the localic restriction of Döring's and Isham's approach is easily overcome by the fact that all elementary topoi are at least ‘internally' locales, elementary theory itself is restricted  in the precise sense that this ‘internal' truth is still essentially propositional, and thus unsuitable to many-body situations where spacetime kinematics itself (read, the rules of logic) might change.

$\ $ \\ \noindent\textit{2.2 Objectivism versus Subjectivism}

\vspace{.2cm}
\noindent
Despite the aforementioned limitations, topos quantum physics shows how we can theorise about the ‘interiority' of physical phenomena in a mathematically intelligible way. And, in fact, even if the aforementioned restrictions apply also to an alternative, ‘covariant'\footnote{The alternative approach is based on the \textit{covariant} lattice of the commutative contexts of $C^*$-algebra [e.g., 53].} approach to topos quantum physics [27], the existence of the latter shows that there exists some level of freedom at least at the level of the articulation of the quantum topos. 

Indeed, aside some technical differences [see Corollaries 2 and 5,  53: 16, 23], the latter approach reflects Bohr's ‘subjectivist' rather than Einstein's ‘objectivist' reading of quantum states: in the contravariant approach a convention is valid only if it applies in all contexts whereas the covariant approach validates a proposition if it holds in at least a single context [53]. In particular, where 
 Döring and Isham thus engage the ‘objectivist' view ‘see[ing] the Kripke--Joyal semantics of the topos [$\sets^{\V(\Ns)}$] as a (physical) Kripke model' [53: 51] and suppose it to provide ‘an observer-inde\-pendent, non-instrumentalist interpretation' [14: 2], the covariant approach proves that, against their belief, the ‘Bohrian', subjectivist view is no more prone to ‘presuppos[ing] a divide between system and observer'. 
 
 In the end, the conflict comes down to what precisely is meant by the notions of ‘localisation' and a situation ‘internal' to a topos. Both views, still representing the \textit{mechanical} point of view over quantum physics, assume the truth ‘internal' to physical systems to be commutative. Temporality is then reduced to the concept of order and negatively articulated as something that is \textit{not} commutative or is ‘indeterminate'. By contrast, we will next illustrate that something remains-there in the second sense of appearing; something that makes even the direction of time, which elementary theory still represents as a global arrow $\underline{\check \Xi} \to \underline{\check \R}$, itself contextual.

\section{Second Quantisation---In Between Bodies}

Elementary topos quantum physics and the first quantisation problem treat quantum phenomena from the perspective of a single, \textit{global concept of locality}. The question of quantities is then expressed by a 1-categorical arrow $\underline{\check \Xi} \to \underline{\check \R}$, but it can only be used to describe single particle systems. By contrast, in the context of many-particle systems we should ask about the symmetries or conjugates that ‘remain there': what is the map $\underline{\check \Xi} \to \underline{\check \R}$ itself equivalent with? 

As one possible answer, Flori [19] uses group theory to address global symmetries of the arrow. Her use of set-theoretic description, however, again delimits the nature of such symmetries. The problem is more conceptual and relates to the way of theorising in mathematics itself: what are the coexisting ways in which the question of quantisation ('daseinisation') may be posed?

Indeed, Richard Feynman's discoveries contest the global divide between an ‘observer' or quantification ($\underline{\check \R}$) and the ‘system' ($\underline{\check \Xi}$). He ran to difficulties while seeking to study many-particle systems propositionally, by statements of the form ‘$\Lambda \in \Delta$'. This fact became apparent as Feynman sought to extend the Lagrangian approach to studying action in the context of indeterminate, stochastic systems. Contrarily to classical mechanics, the Lagrangian energy-regulator itself is then a \textit{stochastic} process, and so are the paths taken by individual, interacting particles.

$\ $

\noindent \textit{3.1 Set-Theoretic Approach: Quantum Field Theories}

\vspace{.2cm}

\noindent
Feynman became synonymous to how controversial quantum physics was about to become. In response to these difficulties, gauge theories and other more advanced field theories emerged as an alternative to the so-called quantum chromo dynamics which, in the way it was expressed at that time, sought to extend  classical quantum theory [43]. 
Instead of starting with individual particles or states,  materiality was regarded as a ‘field' spanning over spacetime, and individual quanta would become identifiable only retrospectively---this identification refers to the so-called \textit{second quantisation} problem. These fields, in turn, are subject to \textit{local symmetries}, that is, the field itself should be unaffected by the change of (also) the local coordinate system.

Quantum electro dynamics provides the easiest example to understand the local coordinate invariance. The field symmetries in this case are ordinary circles (referring to the so-called ‘phase' of a particle), and this simplest theory is also compatible with general relativity. Technically, the coordinate-invariance is directly visible from the path-integral \begin{equation}\left| \int D x \cdot e^{i / \hbar \A} \right|^2 /N\end{equation}
as the norm $|\cdot|$ is unaffected by complex conjugation (or the circle formed by the roots of unity in the complex plane). This theory is implicit to the traditional Maxwell-equations.

However, Yang and Mills [54] were the first to suggest that there are also \textit{non-commutative} local symmetries like the $U(1) \times SU(2)$ symmetry of the electro-weak interaction. 
Symmetries based on such \textit{non-commutative} groups turn out to result in the breaking of the local symmetry and make the theory incompatible with gravity, because the so-called connection field, which is quantised as boson (e.g., photons), is not massless unless it is ‘free', that is, unless there are no interactions\footnote{
Mathematically, field theory describes matter fields as cross sections of a so-called fibre bundle $E \to B$ [5]. The \textit{connection field} associated with that fibre bundle is a form which enables the ‘derivation' of the matter field along the bundle. The connection field in the $U(1)$-invariant theory over $B =\R^{3, 1}$, for example,  
is regulated by Maxwell-equations describing the standard laws of electro dynamics.}. In actuality, indeed, gauge bosons are affected by gravity. 

$\ $ \\ \noindent\textit{3.2 A Categorical Approach}

\vspace{.2cm}

\noindent
To overcome the pitfalls of classical field theory, mathematical physicists have suggested a \textit{categorical} approach to modeling quantum interactions. These interactions are indeed understood as diagrams---in a way not necessarily incompatible with those introduced by Feynman. In particular, it is believed that interactions are too complex processes to be described non-geometrically, by the means of propositional language.

Instead of assuming spacetime kinematics to globally exist, the state space itself is situated or transformed through interactions. For instance, Baez and Lauda [3] accessibly demonstrate how monoidal categories, which are the simplest examples of higher categories, can be used to problematise or ‘localise' the direction of time (a critical issue that has bothered most post-War theorists [e.g., 44]). 
This brings us to the question of the second temporality: \textit{coexistence} of not only objects but also of the relationships between them, including the arrows of time which are opposite to particles and their respective anti-particles. As regards the first quantisation problem ($\underline{\check \Xi} \to \underline{\check \R}$), the direction of time is already implied in the very form of that arrow.

We should rather think that each particle carries its own concept of spacetime, as represented by the state object, and these different concepts come in contact though quantum interactions. This can be found analogous to how Heidegger [25: 383] suggests that the being-there of time is ‘always bound up with some location', reverting Hegel's understanding of ‘being-there' as a determined being (i.e. as represented by the state object). On the level of mathematical theory, however, this idea is problematic. For if there are multiple bodies of spacetime, how can we choose one of them to begin with? To address this crucial question, let us discuss next what mathematics itself says about situations consisting of several bodies.

\section{Coexistence as a Mathematical Problem}
\label{problem}

It is one thing to address interactions and many-body systems mathematically, and another thing for multiple mathematical descriptions of spacetime to coexist. Yet, we argue, there seems to be something that connects these two meanings: quantum field theory as it localises coordinate transformations, and topos theory as it localises the very concept of locality (and spacetime). 
We will begin by discussing this on the side of mathematical coexistence, introducing Alexander Grothendieck's discovery of stacks. In Section 5 we will then discuss their relevance to theories of quantum interaction.

$\ $ \\ \noindent\textit{4.1 Coexisting Universes of Interaction}

\vspace{.2cm}

\noindent
As discussed above, Yang--Mills theory describes quantum interaction and \textit{physical coexistence} as a Lagrangian system, that is, as a fibre bundle with a smooth connection field (i.e. the gauge field). 
Grothendieck [23], in turn, studied how such fibre bundles can \textit{mathematically coexist}, which lead to his discovery of higher categories. Higher categories emerged as a mathematical response to the problem of classifying the so-called fibre spaces. This is the same as asking how physical coexistence is mirrored in the problem of mathematical coexistence. 

The problem of mathematical coexistence is often known as a ‘moduli problem': could the set of all fibre bundles given over a base variety $B$ be itself considered as  a \textit{universal} fibre bundle so that all other bundles would exist as its pull-back constructions. Translating this into the context of physics: could there be a universal system of interactions, that is, could there be a ‘universal' field situating all others. Philosophically, the existence of such a universal system serves as a contemporary version of Hugh Everett's universal wave-function. Otherwise, if a physical system was not universal, how could it recognise its own particularity? 

It is not clear that such universal systems must exist even mathematically, however, which is the precise motive inspiring the shift to categorical mathematics. Set-theoretically such universal systems or entities appear to exist only under special circumstances\footnote{For instance, for the group $G$ being the real line $\R$, the moduli problem is resolved by the so-called projective space $\R\P^{\infty}$. Also, for a topological group $G$, the moduli exists as a classifying $1$-topos consisting of the so-called torsors.}.
Indeed, difficulties arise when the fibre bundle is endowed with a connection field, that is, when the fibres are related to or allowed to ‘interact' with each other. For example, the \textit{universal bundle} of connected $U(1)$-bundles exists neither as a set or as an elementary sheaf-object, but only as the 
\textit{second Deligne complex} 
\begin{equation}\label{circulargerbe} \bar \B U(1) = \Hom(P_1(\cdot), \B U(1)),\end{equation}
which is a ‘stack', that is, a category fibred on groupoids. The theory of stacks generalises ordinary sheaf theory, resulting in 2-categories and 2-topoi instead\footnote{In addition to the objects and arrows (as in a 1-category), a 2-category consists also of the so-called 2-arrows that relate 1-arrows to each other. }. 

The moduli problem is pivotal for understanding the meaning of stacks. Indeed, the local symmetries described by a quantum group $G$ may be conceived as automorphisms ‘internal' to a matter field. These automorphisms are \textit{not} explicated on the level of set-theoretic description of a given fibration, but information about possible conjugate-symmetries (invisible on the level of a single fibre) can only be represented in a categorical setting, which encodes a combination of fibres in relation to each other. 
The symmetries, in turn, cannot be ignored when seeking to incorporate a variety of bundles as a universal one. Namely, each automorphism internal to a given bundle would result in a separate point in the universal bundle, contradicting with the universality condition. As a universal entity needs to encode these ‘internal' automorphisms, it can exist only as a classifying stack. 

$\ $ \\ \noindent\textit{4.2 Higher Topoi}

\vspace{.2cm}
\noindent
To understand the structure of such entities, a $2$-category consisting of stacks is similar to the 2-category of all ordinary categories ($\Cat$). This means that any functor between categories (1-arrow) is intelligible only up to the so-called ‘natural equivalences': it is meaningful to identify a given category only up to the class of similar categories [16], which is why the notion of equivalence is said to ‘weaken' [32: 2]. Similar structures of equivalence are taken as the starting point in the theory of stacks. However, they form more specific class of objects: not only a general 2-category but a 2-\textit{topos}: they are similarly categories giving rise to an ‘internal' concept of truth, but in higher topoi this ‘internal' truth is not propositional but ‘geometric' instead. 

In general, in a higher topos \textit{truth is a relationship} rather than a part (truth was classically interpreted as a part of all possibilities). 
As a consequence, even if it is still possible to talk about a situation ‘internal' to such a topos, the situation is very different: there are no statements in the sense of formal logic but the so-called ‘$n$-geometric' objects instead.  In particular, we cannot think of what is true and false as the opposites or that we could make a point about them.

$\ $

\section{Gerbital Obstruction to Yang--Mills Theory}
\label{gerbe}




Previously we discussed how physical field symmetries prevent us from approaching interactive phenomena by just looking at what happens ‘inside’ different, separate points in spacetime. This is the easiest to understand when spacetime itself is only a single point. Attempting to classify group-actions over a point, Grothendieck discovered that the moduli stack $\Bt G = [\bullet/G]$ is a so-called gerbe, that is, a specific type of a stack (in which any two objects are locally isomorphic). Its fibres are not the action-group $G$ but the classifying topos $BG$ instead. In particular, because the moduli space does not exist as a set, there is something about the structure of that group that is invisible to the situation ‘internal' to its set-theoretic description as a set $G$. 

In effect, symmetries over a single point do not only tell something about that point itself but about the limitations of representing symmetry as a set-theoretic concept, that is, as based on a point-wise description. Following Lawvere's [33: 8] claim, we need a categorical, more ‘general concept' of group-actions. This idea, we will next see, pertains to the second quantisation problem in physics. 

$\ $ \\ \noindent\textit{5.1 Faddeev--Mickelsson Anomaly}

\vspace{.2cm}
\noindent
The explicit fibre-space construction of the matter and connection fields (Section 3) is Cartesian in the sense that the fibres on which the symmtery group acts are perpendicular to spacetime. Quantum field theories attempt to analyse quantum interactions by rearticulating them as combinations of separate particles, that is, by configuring multiple fibre-space constructions together as the so-called Fock-space\footnote{\label{fock} For a connection field $A$, the successful quantisation entails a  decomposition $\Hs = \Hs_+(A) \oplus \Hs_-(A)$ which gives rise to the so called \textit{Fock space} \\ $\F_A  = \bigoplus_{p ,q} \left(\bigwedge^p (\Hs_{+}(A) \otimes \bigwedge^q \bar \Hs_-(A)\right)$
where each term stands for the fields with $p$ particles and $q$ anti-particles.}. This, in turn, would be a Cartesian, direct sum of single-particle systems. However, this renormalisation procedure is obstructed by the so-called Faddeev--Mickelsson operator anomaly [see 18, 39], which stands for local symmetry breaking in quantisation of massless chiral fermions interacting with external gauge potentials. It can be described as a gerbe 
as we will next illustrate following Tähtinen's [51] original construction. It is worth noting, however, that there are actually several equivalent ways to describe the anomaly as a gerbe [26]. 

$\ $ \\ \noindent\textit{5.2 A Sketch of the Construction}

\vspace{.2cm}
\noindent
To sketch out the first construction, gauge formalism operates on fibre bundles defined over indeterminate spacetime. It is not the deterministic, four dimensional one, but instead represented through the operator $^*$-algebra $\As$. If the Hermitean $^*$-structure then extends over the entire bundle (as a morphism $\mathcal{E} \times \mathcal{E} \to \As$), the group of gauge transformations can be defined as the group of unitary transformations\footnote{Formally, the gauge transformations form the group $\Gs(\mathcal{E}) = \{u \in \End_{\As}(\E) \mid u u^{*} = u^{*} u = 1\}$.}. There is a certain renormalisation procedure [39] resulting in a Fock bundle.

The anomaly then occurs when one would like to lift the action of the gauge group $\Grp$ onto  the Fock bundle $\F$---a bundle that represents multiple instances of spacetime, a separate one for each particle\footnote{In technical terms, one would like it to induce a commutative diagram
$$\xymatrix{\F \ar[r]^{\Gamma_A(g)} \ar[d] & \F \ar[d] \\
\A \ar[r]^g & \A}$$ for which $\Gamma_A(g) \hat D_A \Gamma_A^{-1}(g) = \hat D_{A^g}$, where $\hat D_A$ is the second quantised Dirac operator. The existence of such lifting would require $\Gamma_A(e^{iX}) = e^{i}\Gamma_A(X)$ for all $X \in \Lie(\Grp)$ but instead when one moves to the second quantisation, a so called Schwinger term---a $Map(\A, \R)$-valued Lie algebra cocycle of $\Lie(\Grp)$---occurs. }. That lifting is obstructed, however, by the so-called Schwinger obstruction term\footnote{According to Gauss law generators acting on functions $\varphi: \A \to \Hs$, there is a  decomposition $G_A(X) = X + \Ell_X$ with a Lie derivative $\Ell_X$ for $A \in \A$ and $X \in \Lie(\Grp)$ and then $d\Gamma(G_A(X)) = d\Gamma_A(X) + \Ell_X$. This induces an incompatibility with the Lie bracket: 
$[d\Gamma(G_A(X)), d \Gamma(G_A(Y))] = d\Gamma([G_A(X), G_A(Y)]) + c(X, Y; A)$ where $c(X, Y; A)$ is the Schwinger obstruction term.}. This obstruction---the Faddeev--Mickelsson anomaly---is representable by a \textit{gerbe class} $\omega \in H^2([\A/\Grp], \underline S^1)$, where $[\A/\Grp]$ is the quotient stack [see 51: 1084; 52: 42]. 

To translate, it is the ambiguous ‘vacuum level' $\lambda$, indeed, that appears to prevent a field-theoretic resolution of local gauge symmetry. There is no single Fock space but an \textit{entire bundle} of such many-particle constructs, each fibre corresponding to a different energy level. The transformations between these ‘vacuum levels' obstruct the lifting of local symmetry to the Fock bundle\footnote{For two vacuum levels $\lambda$ and $\mu$, given the local neighborhood 
\begin{equation}U_{\lambda \mu} = \{A \in \A \mid \lambda, \mu \notin \Spec (D_A)\} \subset \A,\end{equation}  
with $D_A$ being the Dirac operator, there exists a line bundle $\Det_{\lambda \mu}$ and these local bundles give rise to a bundle gerbe over $\A$ which descends to the moduli space $\A/\G_e$ of equivariant bundles---an entity known to be a smooth, infinite dimensional Fréchet manifold. [42].}.

To summarise our previous discussion, the obstruction to Yang--Mills theory, which is described as a gerbe rather than as a field, seems to require physics to reconsider the kind of a ‘topos' in which physical theory should be expressed: in contrast to sets or sheaves, gerbes occupy a 2-topos instead of 0- or 1-topoi. Each kind of a ‘topos', in turn, implies its own concept of ‘locality', that is, of the structure of topological operations which is not necessarily classical. Therefore, we cannot just say that phenomena are ‘non-local', as implied by those scholars who assume set theory and logic as the proper place of articulating physical theory [e.g., 28, 38], but instead the very concept of locality is ambiguous.

\section{Two Ways for Theorising to End} 
\label{spec_end}

The topos physics perspective suggests us to ask what a ‘final theory'---a ‘closure' of physics like Leibniz would say---could possibly mean. Given how topos theories themselves proliferate (Table 1), the problem of a final theory itself emerges in  ever more complex ways. What is common to all topos perspectives, however, is the fact that this question can be studied in two contrary ways: the possibility of a closure as a part of that topos itself (\textit{internal}\footnote{This ‘internal' geometry, however, is not logical except in the case of elementary theory.}) and as we speculate on the existence of that topos (\textit{external}). 

The first, internal closure is relevant from the point of view of physical systems themselves. 
How could a physical system result in a conclusive description of itself (assuming that any descriptions made by physicists are, indeed, part of the material reality)? In particular, could we say something definite about the kind of phenomena that could be representable even in principle? On the ‘existential' side, by contrast, we are interested in the existence of that topos that could serve as an adequate correlate of physical reality.

$\ $ \\ \noindent\noindent \textit{6.1 The Internal Side: What do Quantum Groups Enable?}

\vspace{.2cm}
\noindent
On the level of the \textit{contents} of quantum theory, the question is essentially about how physical phenomena could result in something reasonable at all. The nature could represent virtually anything (e.g., virtual particles), but we are interested in those representations that can be recognised as its parts and thus be conceived as consequential even in principle. 
\textit{Geometric representation theory} can be viewed as a field studying this question: under what mathematical conditions may local symmetries result in something representable at all. 

Starting with the most obvious case, the circle, the Pontrijagin duality  states that all commutative locally compact groups are recoverable from the corresponding dual groups consisting of their circular characters (continuous homomorphisms $\chi: G \to U(1)$). The circle itself exists as a dual to the group of whole numbers, and thus illustrates the simplest way for physics to support the question of quantifiability. 

The Tannaka--Krein duality extends this duality to compact but \textit{non-commutative} topological groups, which are recovered by the category of representations $\Rep_ {\C}(G)$. This category is commutative and thus well understood, even if $G$ itself is not. However, when non-commutative groups \textit{coexist}, the resulting representation categories are also categorically non-commutative, braided structures\footnote{A braided structure is a bicategorical relation $\gamma_{f, g}: f \otimes g \to g \otimes f$ which satisfies the so called hexagon axiom of associativity, but not the commutativity constraint.}. This concerns the direction of time, for example, as fermions appear to ‘borrow' energy from the corresponding anti-particles that move in the opposite direction of time.

For rigid monoidal categories with both a quantum trace and co-quantum trace, there is, however, the 
Drinfel'd--Jimbo-duality. Ordinary quantum groups like $SU(2)$ and $SU(3)$ satisfy these conditions, thus actualising the possibility that interactions, too, turn representable [see 12]. The geometric Langland's program now continues this quest for a functorial theory of automorphisms, and it has recently been brought in contact with field theory [20].

$\ $ \\ \noindent\textit{6.2 The Existential Side: How are Conditions Possible?}

\vspace{.2cm}
\noindent
As illustrated above, quantum groups emerge as such forms that, if anything, can result in something reasonable. Yet they only make physics \textit{possible}: they are forms that \textit{can} actualise something reasonable, and are thus free to become represented in different positions. But these forms themselves are unable to \textit{necessitate} the exact topos behind spacetime. 

Indeed, if higher theory mirrors many-body systems in the way that the theory itself combines multiple topoi, by assuming that theory itself to exist in a \textit{single} (higher) topos brings us back to the dilemma of the single-many-world problematic, even if the contents of this problem are no longer logical: there is a ‘minimality' condition, a \textit{global} measure, that appears to form the crux of quantum gravity [22]. In the context of this paper, this problem is illustrated by the ambiguous choice of the vacuum level: as discussed above, the vacuum level is, indeed, a global measure, but it cannot be localised in a fashion similar to field theory because the corresponding dilation group $\R_+$ is not compact:  there is no representation-theoretic duality analogous to those discussed above. In particular, there are currently no known ways of combining non-compact and non-commutative theories.

Topos quantum physics provides a conceptual perspective to this problem, addressing the possibility of a closure on the ‘global', existential side. The rather ‘Kantian' question of what permits the emergence of representations is  coupled by the question of the conditions under which representation theory itself is possible.

Döring's and Isham's project is pioneering in this direction, even if their decision to restrict to local theory leaves them short handed (e.g., their discussion of the category of systems as a basis for many-world interpretations). However, 
Butterfield, Döring and Isham believe that this track could help us reformulate theories like the \textit{loop quantum gravity} (LQG), which they believe to be the most promising theory of gravity. The LQG extends Roger Penrose's [42] original work on spin networks\footnote{Spin networks are 1-complexes representing the minimal length problem topologically; these networks then represent the ‘quantum states' of space time itself ('expression') rather than of individual particles ('contents'). The covariant LQG extends extends them to $2$-complexes.} by rearticulating the mass gap problem as a minimal area problem of space-time. 

By ‘quantising' the spacetime itself instead of only particles and interactions, loop quantum gravity could be viewed as a response to a ‘third' quantisation problem. What makes the ‘mass gap' puzzling is the problem of combining these different quantisation problems into a single mathematical theory.  In the light of both the Faddeev--Mickelsson anomaly and the 4-geometric contents of the LQG, it appears that higher categorical methods are needed, and topos quantum physics should reorient its questions accordingly. Indeed, over the past two decades higher geometry has surfaced in physicists' own discourses---especially after Seiberg and Witten [47] suggested the  geometry of spacetime itself to be non-commutative, applying Alain Connes' ideas. 

Of course, it is a possibility that higher geometric thinking could prove out to be redundant. For example, the LQG could prove the extra dimensions assumed by  Yang--Mills theory dispensable, enabling Wilson loops to be used as the basis for nonperturbative quantisation of gravity. However, it would be difficult to prove that the topos within which such a cancellation are expressed is necessarily the classical one. And other than Thomas Thiemann's [50] \textit{Phoenix project}, there is little insight into how that combination should elucidate itself even in principle [but see 1].

\section{Results and Discussion}

Quantum physics can be said to have reverted the classical question of physics: instead of explaining physical phenomena based on a pre-existing quantitative framework (particles), it asks how those quantities (and quantifiability) emerge as a result of nature itself. In this paper we have demonstrated that topos quantum physics, once again, reverts this picture. It poses the question where precisely, that is,  in which exact topos the quantisation thematic itself may intelligibly exist. One crucial open question then is to what extent physical phenomena allow contingency in regard to the choice of that topos. 

This paper has sought to illustrate the relevance of this question in the context of the three quantisation problems of physics.  As the main result of this paper, it appears that the different quantisation problems are associated with different kinds of topoi. As a coarse framework we suggested that it would be helpful to associate (1) quantum mechanics with elementary theory, (2) field theory with the bicategorical constructs, and (3) the ‘third' quantisation of spacetime itself (e.g., LQG) with higher theory, as indicated in Table 1.

At the same time, the question of what makes a theory ‘final' is one of the fundamental problems of physics, mirroring Peter Galison's [21] question of \textit{How Experiments End?} but in the context physical theorising. We argued the question of what makes a theory ‘final' should cohere with the notion of finality internal to its topos. For example, in an elementary topos the concept of finality still plays out as the ‘terminal object', $1$, which then regulates the partitioning of all other objects as a kind of a ‘block universe view' internal to mathematics [cf. 44]. Truth, there, is still defined naturalistically as an ‘incorporeal' relationship: as one contained in the set of all possibilities [cf. 8].

 In a higher topos, by contrast, there is no such a global yardstick. We can surely define truth or finality as a so-called ‘universal full and faithful (op)fibration' [36], but this relationship is not  understandable as an incorporeal property of a single object. In particular, because truth is a relationship and not a subobject, it does not provide a global hierarchy of truth-values\footnote{For an object to be ‘contained' in another one is considered as being a unique, unambiguous relationship---an ‘incorporeal' property of the object $\Omega$ itself. Such an inclusion-relationship can be viewed as a special case of the more general definition of the arrow ‘true' as a fibration. What makes the incorporeal definition of an elementary topos specific is that objects in it form a global hierarchy, while those in a higher category do not.} that would give rise to an ‘internal' logic or quantum algebra like in the case of single-particle systems. 

$\ $ \\ \noindent\textit{7.1 Conclusions}

\vspace{.2cm}
\noindent
We first followed Wolters' [53] comparison of the two elementary approaches to the first quantisation problem, illustrating the connotations between the classical debate (subjectivist and objectivist interpretations) and elementary theory. Second, we discussed anomalies related to theories of gravity and interaction. We then concluded that the possibility of resolving the problem of a final theory in the classical sense of propositional logic would require the higher geometries associated with interaction and loop quantum gravity to be cancelled out. By contrast, if higher theory is necessary for understanding physical reality, its implications to philosophy are  foundational. Not only would physical phenomena be ‘nonlocal', like most quantum philosophers now argue (e.g., Holman, 2014), but there would not even exist a global concept of locality. 

We do not argue that science necessarily ‘advances' in the direction indicated in Table 1 [cf. 31], however, as the concept of higher theory is itself ambiguous [cf. 34]. Nevertheless, we can contest the view that physical reality makes mathematics as such ‘fail' (cf. Lisi and Weatherall, 2010). Rather, mathematics itself is  a temporal, evolving process [cf. 32], and given how Yuri Manin (personal communication) reflects higher theory as ‘post-modern' mathematics, Sokal’s mischievous proposal for a postmodern geometry of quantum gravity is rather ironic. 

%

\newpage

\section*{References}

\setlength{\parindent}{-7mm} 
\setlength{\parskip}{.2mm}
\setlength{\leftskip}{7mm}

$ $

[1]\hspace{2mm} Alexander, S., Marciano A., Tacchi, R.A.: Towards a Loop Quantum Gravity and Yang--Mills unification. Physics Letters B 716(2): 330--333 (2012)

[2]\hspace{2mm} Awodey, S: Structure in Mathematics and Logic: A Categorical Perspective. Philosophia Mathematica 4 (3): 209--237 (1996)

[3]\hspace{2mm} Baez, J.C., Lauda, A.D.: A Prehistory of $n$-Categorical Physics. In: Halvorson, H. (ed.) Deep Beauty: Mathematical Innovation and Research for Underlying Intelligibility in the Quantum World, pp. 13--128. Cambridge: Cambridge University Press (2011) 

[4]\hspace{2mm} Bain, J.: Three principles of quantum gravity in the condensed matter approach. Studies in History and Philosophy of Modern Physics 46: 154--163 (2014)

[5]\hspace{2mm} Bernstein,  H.J.,  Phillips, A. V.: Fibre Bundles and Quantum Theory. Scientific American 245: 94--109 (1981)

[6]\hspace{2mm} Bohr, N.: Discussion with Einstein on Epistemological Problems in Atomic Physics. In: Einstein, A. (ed.) Philosopher-Scientist, pp. 201--241. La Salle: Open Court (1949)

[7]\hspace{2mm} Bowker, G.C., Star, S.L.: Sorting Things Out: Classification and its
Consequences. Cambridge: MIT Press (1999)

[8]\hspace{2mm} Brown, J.R.: Platonism, naturalism, and mathematical knowledge. New York:  Routledge (2013)

[9]\hspace{2mm} Cohen, P.J.: The Independence of the Continuum Hypothesis. Proc. Natl. Acad. Sci. USA 50(6): 1143--1148 (1963)

[10] Cohen, P.J.: The Independence of the Continuum Hypothesis II. Proc. Natl. Acad. Sci. USA 51(1): 105--110 (1964)

[11] Crowther, K., Rickles, D.: Introduction: Principles of quantum gravity. Studies in History and Philosophy of Modern Physics 46: 135--141 (2014)

[12] Drinfel'd, V.G.: Quantum groups. Proc. In: Gleason, A. (ed.) Intl. Congress of Mathematicians, pp. 798--820. Berkeley: Berkeley University (1987)

[13] Döring, A., Isham, C.: A Topos Foundation for Theories of Physics:
II. Daseinisation and the Liberation of Quantum Theory. Journal of Mathematical Physics 49(5): 053516 (2008)

[14] Döring, A., Isham, C.: Classical and Quantum Probabilities as Truth Values. Journal of Mathematical Physics 53(3): 032101 (2011a)

[15] Döring, A., Isham, C.: ‘What is a Thing?': Topos Theory in the Foundations of Physics. In Coecke, B. (ed.) New Structures for Physics, pp. 753--937. Berlin \& Heidelberg: Springer (2011b)

[16] Eilenberg, S., Mac Lane, S.: General theory of natural equivalences. Transactions of the AMS 58: 231--294 (1945)

[17] Everett, H.: The theory of the universal wavefunction. In: DeWitt, B., Graham, N. (eds.) The Many-Worlds Interpretation of Quantum Mechanics, pp. 3--140. Princeton: Princeton University Press (1973)

[18] Faddeev, L.: Operator anomaly for the Gauss law. Physics Letters 145B: 81--84 (1984)

[19] Flori, C.: Group Action in Topos Quantum Physics. Journal of Mathematical Physics 54(3): 032106-52 (2013)

[20] Frenkel, E., Witten, E.: Geometric endoscopy and mirror symmetry. Available via arXiv. http://arxiv.org/pdf/0710.5939. Cited 10 Jan 2017 (2007)

[21] Galison, P.: How Experiments End. Chicago: University of Chicago Press (1987)

[22] Garay, L.J.: Quantum gravity and minimum length. International Journal of Modern Physics A 10(2): 145--165 (1995)

[23] Grothendieck, A.: General Theory of Fibre Spaces with Structure Sheaf. National Science Foundation Research Project on Geometry of Function Spaces. Report 4. Lawvrence: University of Kansas (1958)

[24] Halvorson, H. (ed.): Deep Beauty. Understanding the Quantum World through Mathematical Innovation. Cambridge: Cambridge University Press (2011)

[25] Heidegger, M.: Being and Time. Stambaugh J. (trans.). New York: State University of New York Press (1953)


[26] Hekmati, P.M.K.,  Murray, D.S., Vozzo, R.F.: The Faddeev--Mickelsson--Shatashvili-Anomaly and Lifting Bundle Gerbes. Communications in Mathematical Physics 319(2): 379--393 (2013)

[27] Heunen, C., Landsman, N., Spitters, B.: Bohrification of operator algebras and quantum logic. Synthese 186(3): 719--752 (2012)

[28] Holman, M.: Foundations of quantum gravity: The role of principles grounded in empirical reality. Studies in History and Philosophy of Modern Physics 46: 142--153 (2014)

[29] Jaffe, A., Witten, E. Quantum Yang--Mills Theory. Available via Clay Mathematical Institute. \\ http://www.claymath.org/sites /default/files/yangmills.pdf. Cited 2 Jun 2016 (2016)

[30] Johnstone, P.T.: \textit{Topos Theory}. London: Academic Press (1977) 

[31] Kitcher, P.: The Advancement of Science: Science without Legend, Objectivity without Illusions. Oxford: Oxford University Press (1995)

[32] Landry, E., Marquis, J-P.: Categories in Context: Historical, Foundational, and Philosophical. Philosophia Mathematica 13: 1--43 (2005)

[33] Lawvere, W.: Functorial Semantics of Algebraic Theories. Reprints in Theory and Applications of Categories 5: 1--121 (2004)   

[34] Leinster, T.: Higher Operads, Higher Categories. London Mathematical Society Lecture Notes Series. Cambridge: Cambridge University Press (2003) 

[35] Lisi, A.G., Weatherall, J.O.: A Geometric Theory of Everything. Deep down, the Particles and Forces of the Universe are a Manifestation of Exquisite Geometry. Scientific American 303(6): 54--61 (2010)

[36] Lurie, J.: Higher Topos Theory. Princeton: Princeton University Press (2009)

[37] Marquis, J-P.:From a Geometrical Point of View. A Study of the History and Philosophy of Category Theory. New York: Springer (2009)

[38] Mattingly, J.: Unprincipled microgravity. Studies in History and Philosophy of  Modern Physics 46: 179--185 (2014)

[39] Mickelsson, J.: Regularisation of current algebra. In Constraint theory and quantisation methods, Mentepulciano. River Edge: World Science Publishers (1993)

[40] Mills, M.A.: Identity versus determinism:  Émile Meyerson's neo-Kantian interpretation of the quantum theory. Studies in History and Philosophy of Modern Physics 47: 33--49 (2014)

[41] Pelc, O., Horwitz, L.P.: Generalisation of the Coleman--Mandula Theorem to Higher Dimension. Journal of Mathematical Physics 38(1): 139--172 (1997)

[42] Penrose, R.: Applications of negative dimensional tensors. In: Welsh, D.J.A. (ed.) Combinatorial mathematics and its applications,  pp. 221--244. New York: Academic Press (1971)

[43] Pickering, A.: \textit{Constructing Quarks: A Sociological History of Particle Physics}. Chicago: University of Chicago Press (1984) 

[44] Price, H.: Time's Arrow \& Archimedes' Point. New Directions for the Physics of Times. New York \& Oxford: Oxford University Press (1996)

[45] Sakata. S.: The Theory of the Interaction of Elementary Particles I. The Method of the Theory of Elementary Particles.Progress of Theoretical Physics 2(3): 145--150 (1947)

[46] Schreiber, U.: Differential cohomology in a cohesive infinity-topos. Available via arXiv. http://arxiv.org/pdf/1310.7930. Cited 10 Jan 2017 (2013)

[47] Seiberg, N., Witten, E.: String theory and noncommutative geometry. Journal of High Energy Physics 09: 032 (1999)

[48] Shapiro, S.: Categories, structures, and the Frege-Hilbert controversy: The status of meta-mathematics. Philosophia Mathematica 13(1): 61--77 (2005)

[49] Smeenk, C.: The Elusive Higgs Mechanism. Philosophy of Science 73(5):  487--499 (2006)


[50] Thiemann, T.: The Phoenix Project: Master Constraint Programme for Loop Quantum Gravity. Classical and Quantum Gravity, 23(7): 2211--2247 (2006)

[51] Tähtinen, V.: Anomalies in gauge theory and gerbes over quotient stacks. Journal of Geometry and Physics 58: 1080--1100 (2008)

[52] Tähtinen, V.: On the Geometry of Infinite-Dimensional Grasmmannian Manifolds and Gauge Theory. Ph.D. thesis, University of Helsinki (2010)

[53] Wolters, S.A.M.: A Comparison of Two Topos-Theoretic Approaches to Quantum Theory. Communications in Mathematical Physics 317(1): 3--53 (2013)

[54] Yang, C.N., Mills, R.: Conservation of isotopic
spin gauge invariance. Physical Review 96: 191 (1954)

[55] Zahar, E.: Meyerson's ‘Relativistic Deduction': Einstein versus Hegel. British Journal for the Philosophy of Science 38(1): 93--106 (1987)

\end{document}